\documentclass[pra,superscriptaddress, twocolumn]{revtex4-2}

\usepackage{amsmath}
\usepackage{amsfonts}
\usepackage{graphicx}
\usepackage{nicefrac}
\usepackage{dcolumn}
\usepackage{braket} 
\usepackage{svg}
\usepackage{float}
\usepackage{hyperref}
\usepackage{caption}
\usepackage{subcaption}
\makeatother

\begin{document}

\title{Numerical modeling for trapped-ion thermometry using dark resonances}

\author{Muriel Bonetto}
\affiliation{Universidad de Buenos Aires, Facultad de Ciencias Exactas y Naturales, Departamento de F\'isica,
Laboratorio de Iones y \'Atomos Fr\'ios, Pabell\'on 1, Ciudad Universitaria, 1428 Buenos Aires, Argentina}
\affiliation{CONICET - Universidad de Buenos Aires, Instituto de F\'isica de Buenos Aires (IFIBA), Pabell\'on 1,
Ciudad Universitaria, 1428 Buenos Aires, Argentina}
\author{Nicol\'as Adri\'an Nu\~nez Barreto}
\affiliation{Universidad de Buenos Aires, Facultad de Ciencias Exactas y Naturales, Departamento de F\'isica,
Laboratorio de Iones y \'Atomos Fr\'ios, Pabell\'on 1, Ciudad Universitaria, 1428 Buenos Aires, Argentina}
\affiliation{CONICET - Universidad de Buenos Aires, Instituto de F\'isica de Buenos Aires (IFIBA), Pabell\'on 1,
Ciudad Universitaria, 1428 Buenos Aires, Argentina}

\author{Christian Tom\'as Schmiegelow}
\affiliation{Universidad de Buenos Aires, Facultad de Ciencias Exactas y Naturales, Departamento de F\'isica,
Laboratorio de Iones y \'Atomos Fr\'ios, Pabell\'on 1, Ciudad Universitaria, 1428 Buenos Aires, Argentina}
\affiliation{CONICET - Universidad de Buenos Aires, Instituto de F\'isica de Buenos Aires (IFIBA), Pabell\'on 1,
Ciudad Universitaria, 1428 Buenos Aires, Argentina}
\author{Cecilia Cormick}
\affiliation{Universidad de la República, Facultad de Ingeniería, Instituto de F\'isica, Julio Herrera y Reissig 565, 11300 Montevideo, Uruguay}
\affiliation{Instituto de F\'isica Enrique Gaviola, CONICET and Universidad Nacional de C\'ordoba,
Ciudad Universitaria, X5016LAE, C\'ordoba, Argentina}

\begin{abstract}
    The simulation of vibrational energy transport and quantum thermodynamics with trapped ions requires good methods for the estimation of temperatures. One valuable tool for this purpose is based on the fit of dark resonances in the fluorescence spectrum. However, this fit demands numerical simulations of the coupled electronic-motional dynamics which usually involve a trade-off between accuracy and speed. Here, we discuss several techniques with simplified dynamical equations for the simulation of the spectrum of a trapped ion that undergoes thermal motion, identifying the advantages and limitations of each method. We start with a three-level model to provide a better insight into the approximations involved, and then move on to tackle the experimentally relevant case of an eight-level calcium ion. We observe that mimicking the effect of thermal motion by means of additional dephasing is computationally very convenient, but can lead to significant errors in the estimation of the temperature. Nevertheless, this can be counteracted by a proper calibration, supporting the use of dark resonances as a practical thermometer.
\end{abstract}

\date{\today}

\maketitle

\section{Introduction}

Trapped ions constitute a versatile platform for exploring thermodynamic processes at the microscopic scale. Their ability to form various crystalline structures, such as linear chains, zigzag configurations, or more complex geometries, makes them ideal for investigating heat transport~\cite{Freitas2016,bermudez2013,Ruiz2014,ruiz2019,timm2023}, out-of-equilibrium thermodynamics~\cite{Ramm2014,Onishchenko2024,Meir2018}, and thermal engines~\cite{von2019spin, Rossnagel2014}. For example, these systems can exhibit transitions between ballistic and diffusive transport regimes, offering valuable insight into the emergence of Fourier’s law at microscopic scales~\cite{Freitas2016,Ruiz2014}. The exploration of such phenomena requires accurate techniques for local thermometry. 

Several methods have been developed for temperature determination in trapped ions. For temperatures above a few Kelvin, spatial spreading of the ions can be used effectively~\cite{neuhauser1980,norton2011,boldin2018,Mao2022}. 
For lower temperatures, in the sub-Kelvin regime, Doppler broadening of spectral lines offers temperature estimates constrained by the linewidth of the selected transition, which typically lies in the 10 MHz range for dipole-allowed transitions~\cite{wineland1978,wesenberg2007,epstein2007}.
For even lower temperatures, in the micro-Kelvin range, sharper spectral features such as motional sidebands in quadrupole or Raman transitions 
enable accurate measurements of the thermal population of each vibrational mode. However, they generally do not provide information about local properties~\cite{wineland1987,monroe1995,ivanov2019,Vybornyi2023}.  

A particularly promising ion thermometry approach involves the use of Coherent Population Trapping (CPT), also known as Dark Resonance Spectroscopy~\cite{rossnagel2015fast,peters2012,Tugaye2019}. This method, which relies on a two-photon process, leverages narrow spectral lines to measure temperatures from a few milli-Kelvin down to the micro-Kelvin range. The appeal of this technique lies in its direct connection to local degrees of freedom, as opposed to global measurements of normal mode populations, while still being suitable for low-temperature regimes where quantum phenomena are more pronounced.

Dark resonances manifest in spectra as  sharp reductions in fluorescence with a width and depth determined mainly by the intensity of the lasers and ultimately by the coherence of the interacting laser-atom system. Thermal motion, however, leads to a Doppler shift of the laser fields, broadening and blurring these resonances. This characteristic allows using CPT spectra as thermometers. By fitting such spectral features one can obtain an estimation of the temperature. 

An accurate temperature measurement from a CPT spectrum requires computationally intensive fits of experimental data to theoretical spectra. As a full simulation of the coupled dynamics of all electronic levels and motional states is too costly to be practical, more economical approximations are commonly used. As we show in this paper, while these approximations make calculations more manageable, they also introduce errors, causing different numerical methods to yield varying results.

In this work, we analyze and compare various techniques that can be used to estimate the temperature from a CPT spectrum. In doing so, we determine the quality and regime of applicability of different methods in their ability to model the effect of thermal motion and to extract reliable temperatures from CPT spectra in an accurate and economical fashion. We also provide practical recipes for calibration and use of these methods. For this study, we will assume that the motional state is thermal and can be described semiclassically. The exploration of non-thermal and quantum-mechanical regimes of the motion is beyond the scope of the present analysis. Also, for simplicity, we focus on the case of a single trapped ion, but our results can be extended to more ions, as they depend on the local velocity of each ion.

To model the spectra, we consider the following approaches, which are detailed in section \ref{sec:3levels}: A) the ``oscillatory shift approximation'',  where the ion motion leads to an oscillatory Doppler shift in the electronic evolution, B) the ``instantaneous relaxation approximation'', which neglects the time scale associated with the electronic relaxation; C)~the ``sidebands approximation'', where the internal state is assumed to be periodic and the coupled equations for the different Fourier components are solved; D)~the ``effective dephasing approximation'', in which the thermal broadening of the spectral lines is mimicked by dephasing.  Each of these methods has advantages and problems, and each has a certain regime of applicability, which we will discuss in detail.

Our benchmark will be given by approach A, consisting of the simulation of the electronic dynamics with a time-dependent Doppler shift induced by the atomic motion. We numerically integrate the resulting time evolution of the optical equations until the system reaches the stationary state. By temporally averaging the excited-state population, we estimate the fluorescence of the ion.
Unfortunately, the resulting calculation is too demanding to be practical for spectral fits; instead, it can be used to ``calibrate'' the computationally more affordable methods. 

The work is organized as follows: In Section \ref{sec:thermometry} we briefly describe our model for the thermal motion of a trapped ion and introduce the basic concepts regarding temperature estimation based on dark-resonance spectra. In Section \ref{sec:3levels}, we address the simpler case of a three-level system. We then proceed to the experimentally relevant case of the full eight-level system corresponding to trapped calcium in Section \ref{sec: 8 levels}. Finally, in Section \ref{sec:conclusions} we  summarize our results.

\section{Thermometry based on dark-resonance spectra}
\label{sec:thermometry}

\begin{figure}[ht!]
    \centering
    \includegraphics[width = \columnwidth]{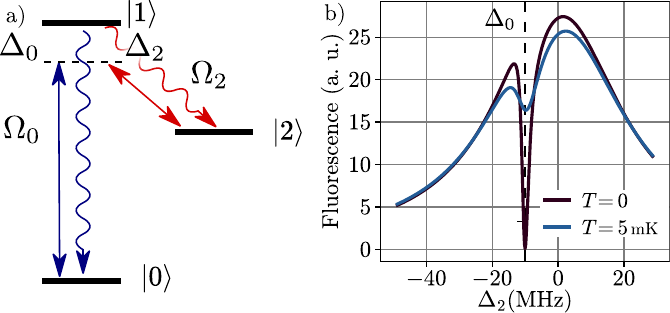}
    \caption{a) Schematic representation of a three-level atom in a $\Lambda$-type configuration. $\Omega_0$ and $\Omega_2$ are the Rabi frequencies of the transitions while $\Delta_0$ and $\Delta_2$ represent the detunings of the respective lasers. b) Example of the CPT spectrum for a three-level system for temperatures of  0 and 5~mK in violet and blue respectively.}
    \label{fig:sketch}
\end{figure}

Our atomic system of interest is such that the relevant electronic levels can be described as a $\Lambda$-type structure, as shown in Fig.~\ref{fig:sketch}~(a). For definiteness, we focus on the case of a trapped calcium ion, where the $\Lambda$-type model is formed by the states $S_{1/2}\leftrightarrow P_{1/2}\leftrightarrow D_{3/2}$ (ground-excited-metastable). This system is driven by a 397~nm laser connecting the levels $S_{1/2}$ and $P_{1/2}$, and a 866~nm laser connecting the levels $D_{3/2}$ and $ P_{1/2}$. Naturally, the ideas presented here are not restricted to the calcium ion and can be applied to other atomic species or laser configurations that exhibit a $\Lambda$-type structure. 

Dark-resonance spectra, such as the one sketched in Fig.\ref{fig:sketch}(b), are obtained by scanning the detuning $\Delta_2$ of the 866~nm laser while keeping the 397~nm laser fixed at a red detuning $\Delta_0$ of approximately half the natural linewidth. The atomic fluorescence is proportional to the population of the excited $P_{1/2}$ state. The resulting spectrum shows a broad feature, whose width is mainly determined by the lifetime of the $P_{1/2}$ level, with a sharp dip occurring when the detuning of the scanning laser, $\Delta_2$, matches that of the fixed laser, $\Delta_0$. As the temperature increases, this dip becomes shallower and broader, as illustrated  for 5~mK in the blue curve of Fig.\ref{fig:sketch}(b).

\subsection{Dark resonances}

For simplicity, we first restrict to only three states, that is, we neglect the effects of degeneracies in the levels illustrated in Fig.~\ref{fig:sketch}. Realistic models with more sub-levels display similar phenomena, as will be discussed in Section~\ref{sec: 8 levels}. Here, we assume that levels $\ket{0}$ and $\ket{2}$ are stable and have dipole-allowed transitions to an excited level $\ket{1}$,  from which the atom can spontaneously decay back to the stable states. Each transition is driven with a Rabi frequency $\Omega_{0/2}$ and a detuning $\Delta_{0/2}$ from the excited state.

If atomic motion is neglected, when the laser detunings are equal there is a coherent superposition of levels $\ket{0}$ and $\ket{2}$ that is not coupled with the light fields. Spontaneous decay eventually brings the atomic population into this state,  so that the atom stops fluorescing. This coherent superposition is called a ``dark state'', corresponding to the coherent population trapping phenomenon described before.

In terms of atomic spectra, when one of the laser fields has a fixed frequency but the other frequency is varied, the existence of a dark state manifests as an abrupt decay in atomic fluorescence, as shown in Fig.~\ref{fig:sketch}~(b). This kind of measurement is called a ``CPT spectrum'' and the decay in emission is a ``dark resonance''.
For an atom at rest, the width and depth of this spectral feature are determined by the linewidths and intensities of the lasers and by the atomic linewidth. When the laser fields are such that the motion induces a different Doppler shift for each transition, the resonances also depend on the velocity $\vec{v}$ of the ion and the wavevectors of the lasers $\vec{k}_{0/2}$. This is because the difference in effective detunings is then given by:
\begin{equation}
    \delta_0 -\delta_2 = \Delta_0 -\Delta_2 - (\vec k_0 - \vec k_2) \cdot \vec v \,.
\end{equation}
Here we take the sign convention for the detuning at rest $\Delta_j = \omega_j^{(l)} - \omega_j^{(0)}$ where $\omega_j^{(0)}$ is the atomic transition frequency and $\omega_j^{(l)}$ is the laser frequency. The fact that dark resonances are sensitive to motion makes CPT spectroscopy a useful resource for atomic thermometry, as will be discussed below.

\subsection{Semiclassical model for the thermal motion of a trapped ion}

We will address the characterization of temperature for a single ion in a radiofrequency trap. Before proceeding with the modeling, a few observations are in order. 

For an ergodic and fully thermalized system, temperature is a measure of the mean kinetic energy. However, in trapped ion systems, the kinetic energy may vary across different degrees of freedom, leading to direction-dependent effective temperatures. In general, one will have to take into account that the temperature that one extracts from a spectrum will depend on the relative geometrical configuration of the lasers and the ion trap. 

Another relevant issue is the fact that under typical operation conditions, the motion of the ion can be split into two components: an oscillation with small amplitude at the radiofrequency of the trapping field, and a slower motion with larger amplitude that can be described as an oscillation in an effective static harmonic potential. The former is usually called ``micromotion'' while the latter is referred to as ``secular motion''~\cite{leibfried2003quantum}. The two kinds of motion have very different properties: most importantly, micromotion is forced motion driven by the trap's oscillatory field. Its value is mainly determined by the ion's equilibrium position and not by thermal effects. In contrast, secular motion is the motion of the ion in the harmonic pseudopotential of the trap and can therefore be associated with the ion's temperature.

The two motional components also have a different impact on the fluorescence spectrum. The frequency of the micromotion is usually higher than the width of the dark resonances. As a result, micromotion generally leads to a frequency-modulated spectrum, with the appearance of echoes and with a corresponding reduction in the depth of these resonances without significant change in their width~\cite{nunez2024dark}. Secular motion, on the other hand, takes place at lower frequencies and results in a broadening of the dark resonances~\cite{rossnagel2015fast}. For simplicity and because in many cases micromotion can be strongly reduced by proper compensation of the trap's potentials, in the following we will neglect micromotion and focus on the spectral effect of the secular motion alone.  

Before tackling the three-dimensional case, we will further simplify our calculation by assuming that the ion's motion is one-dimensional. This approximation is equivalent to neglecting the motion in two of the trap directions, or to assuming that one of the trap axes coincides with the propagation direction of the laser fields used to obtain the CPT spectrum. 

We will also restrict to the case where a semiclassical description of the secular motion is valid. Each instance of the experiment corresponds to a situation in which the atomic position oscillates at the secular frequency $\omega$ with a given amplitude $A$ such that
\begin{equation}
    x(t) = A \sin{(\omega t + \varphi)}
    \label{eq: ion motion}
\end{equation}
with $\varphi$ a random phase. The ion velocity is then
\begin{equation}
    v(t) = A \omega \cos(\omega t+\varphi).  
\end{equation}
For an ion moving at a given velocity, the Doppler effect causes an effective detuning for each laser-driven transition given by 
\begin{equation}
    \delta_j(t) = \Delta_j - k_j v(t) = \Delta_j - k_j A\omega \cos(\omega t + \phi).
    \label{eq:Doppler}
\end{equation}
For our simulations of the electronic dynamics, the motion of the ion will be included as an external harmonic driving at the secular frequency. This time-dependent shift in the effective detunings causes a corresponding shift in the spectral position of the dark resonance and will thus be the basis to infer a temperature from the spectrum.
 
Throughout the following sections, we consider a thermal distribution, so that the kinetic energy of the particle is a random variable that follows a Maxwell-Boltzmann distribution. In our semiclassical treatment of the motion, the amplitude $A$ is related to the mechanical energy~$E$ by $E = m(A \omega)^2/2$. Since we are interested in experiments collecting light for time intervals typically longer than the period of the oscillation, the spectrum corresponding to a given mechanical energy can be obtained by averaging over time. Then, the results corresponding to a thermal state are calculated by averaging over $E$ with a Boltzmann weight. 

We will neglect the backaction of the electronic degrees of freedom on the motion, corresponding to the momentum change upon absorption and emission of photons. This is a reasonable approximation within the semiclassical regime, where typical values of momentum are much larger than the recoil kick, so that the time scale of the atomic motion is fast compared with the time scale over which effects due to these kicks become observable.

\subsection{Temperature estimation from dark-resonance spectra}

As discussed above, Doppler shifts associated with an atomic velocity $\vec{v}$ lead to a change in the position of the dark resonance. Thermal motion in a trap does not correspond to a fixed shift, but rather to an oscillating shift, which has to be averaged over several realizations. As a result, dark resonances are broadened and blurred by the motion. For high enough temperature, the resonances tend to disappear completely. For low temperatures, the effect of the motion tends to be masked by other effects, such as laser linewidth, power broadening, or off-resonant transitions. For intermediate scenarios, the analysis of the dark resonances allows for an estimation of the temperatures~\cite{rossnagel2015fast,Tugaye2019}. 
However, the extraction of a temperature value from an atomic spectrum requires a theoretical model to map changes in the spectrum to changes in temperatures. One possible approach is to fit the measured spectrum to the full model. This is a nontrivial step as an accurate model has the drawback of being exceedingly time-consuming and usually has many free parameters which can make the fit impractical or unreliable. On the other hand, some approximations that render the numerical simulation much faster are not always valid or accurate. 

To include the temperature of the ion in the numerical calculation of the spectrum we follow four different methods. We compare the results of these procedures and discuss the applicability of each one. In the next section we describe each simulation strategy for the case of the three-level system and in the following one we generalize our results to the more complex eight-level system.

\section{Spectrum of a three-level system}
\label{sec:3levels}

 We first discuss the theoretical model of the electronic dynamics. We consider the system described in Fig.\ref{fig:sketch}, as detailed before. Under the rotating wave approximation and in the rotating frame, the electronic dynamics are described by the Hamiltonian

\begin{equation}
H=\hbar\left(\begin{array}{ccc}
\Delta_0 & \frac{\Omega_{0}}{2} & 0 \\
\frac{\Omega_{0}}{2} & 0 & \frac{\Omega_{2}}{2} \\
0 & \frac{\Omega_{2}}{2} & \Delta_2
\end{array}\right)
\label{ec: hamiltonian rwa}
\end{equation}
written in the basis $\{\ket{0}, \ket{1}, \ket{2}\}$. 

To include spontaneous emission and laser imperfections, we add a dissipative Lindblad superoperator to the system's master equation for the density matrix of the system $\rho$ as
\begin{equation}
    \frac{d\rho}{dt}  = -\frac{i}{\hbar} \left[H, \rho \right] + \mathcal{L} \rho\ = \mathcal{M} \rho.
    \label{ec: ecuacion maestra}
\end{equation}
Here $\mathcal{M}$ is the superoperator generating the complete, non-unitary evolution, with dissipative terms given by
\begin{equation}
        \mathcal{L} \rho= \sum_{\alpha=l,d}\,\sum_{j=0,2} \frac{1}{2} \left[C_{\alpha j} \rho, C_{\alpha j}^{\dagger} \right] + {\rm H.c.}
        \label{eq:non-unitary}
\end{equation}
In this expression, $j=0,2$ indicates a given lower state, while $d$ stands for ``decay'' and $l$ for ``linewidth''. The four jump operators $C_{\alpha j}$ are the following:
\begin{equation}
\begin{split}
    C_{dj} &= \sqrt{\Gamma_{dj}} ~\ket{j}\bra{1}\\
    C_{lj} &= \sqrt{2\Gamma_{lj}} \ket{j}\bra{j}
    \label{eq: jump operators}
\end{split}
\end{equation}
with $\Gamma_{dj}$ the spontaneous decay rate from the excited state to the lower level $\ket{j}$, and $\Gamma_{lj}$ the dephasing rates due to the linewidths of the respective lasers.

The master equation can be solved to find the dynamics of the density matrix and the evolution of the ion's fluorescence, which is proportional to the population of the excited state, $\rho_{11}$. The atomic motion leads to an oscillating variation of the detunings resulting from the Doppler shift. In general, the motion can also affect the electronic state through a spatial dependence of the Rabi frequencies. We will not consider this aspect since we assume that the laser field can be treated as a plane wave with constant amplitude and polarization.

\subsection{Oscillatory shift approximation}

The most accurate and time-consuming method we consider, which we call ``oscillatory shift approximation'' and use to benchmark the rest, consists of numerically implementing the Optical Bloch Simulations (OBS) for the time-dependent evolution of the electronic state. 
To calculate the fluorescence from the simulations, we introduce the oscillatory time dependence of the laser detunings, as given by Eq.~\eqref{eq:Doppler}, in the Hamiltonian of Eq.~\eqref{ec: ecuacion maestra} and find the density matrix as a function of time. For each value $E$ of the mechanical energy, the fluorescence is proportional to the time average of the excited state population, $\overline{\rho}_{11}(E)$, after a sufficiently long waiting time so that the numerical solution is approximately stationary.

If one assumes that the ion is in a thermal state in the semiclassical regime, the fluorescence for each set of parameters can be calculated as the weighted average
\begin{equation}
    \mathcal{F}(T) = F_0\, Z^{-1} \int_0^\infty  \overline{\rho}_{11}(E) e^{-{E}/{ k_B T} } dE
    \label{eq: energy average}
\end{equation}
with $k_B$ the Boltzmann constant, $Z$ the partition function and $F_0$ a constant that depends on the light collection efficiency. The desired spectrum is obtained by sweeping over $\Delta_2$ while keeping $\Delta_0$ fixed. 

\begin{figure}[ht!]
\includegraphics[width = \columnwidth]{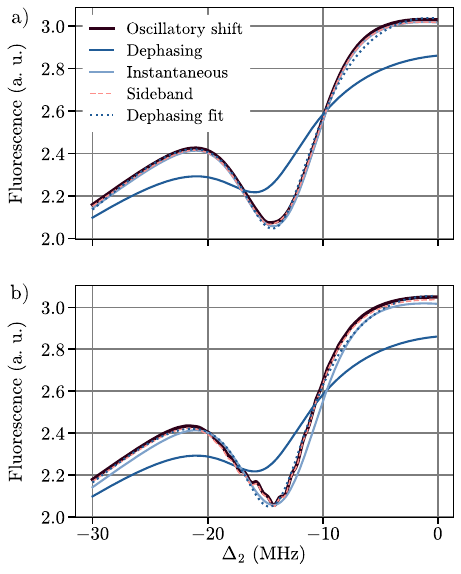}
    \caption{Comparison of three-level spectra calculated with the four methods described, for $T=20$ mK and assuming $^{40}$Ca$^+$ ions driven at the dipolar transitions at 397 nm and 866 nm. We show plots for two different secular frequencies, assuming motion in one dimension. In a) the trap frequency is $\omega = 2\pi\times 0.1$~MHz, while in b) it is $\omega = 2\pi\times 1$~MHz. The dashed blue line is the spectrum calculated by including thermal effects as dephasing with the temperature that yields the best fit, which is $T = 10.8$ mK in a), and $T = 9.6$ mK in b). For both figures, $\Omega_0 = 0.3 \Gamma_{10}$, $\Omega_2 = 3 \Gamma_{12}$ and $\Delta_0 = - 2 \pi \times 15 $ MHz.\label{fig:three levels}}
\end{figure}

In Fig.~\ref{fig:three levels} we show different curves calculated by the methods we consider in this work. The chosen parameters generate, for $T=0$, a dark resonance with a width of $\approx2$~MHz.  
We set $T=20$~mK and plot results for two different trap frequencies, $\omega=2\pi \times 0.1$~MHz in Fig.~\ref{fig:three levels}~a) and $\omega=2\pi \times 1$~MHz in Fig.~\ref{fig:three levels}~b). In both subfigures, the black solid curves represent the calculated spectrum using the oscillatory shift approximation. We observe in a) that when the trap frequency is much lower than the zero-temperature width of the dark resonance, the spectrum appears smooth. In contrast, in b), when the trap frequency is comparable with the scale of the electronic dynamics, we observe small wiggles in the spectrum. 

As a reference of the computational resources needed for this method, we report the simulation time and system used. Simulating one of the curves with this approach on one core of a Dell PowerEdge R720xd takes 20 minutes for a three-level system. On the other hand, it takes over 20 hs running on 32 cores for an eight-level system.  In the following subsections, we will present different approximations that reduce computational cost and discuss how well they reproduce these results.

\subsection{Instantaneous relaxation approximation}
\label{subsec:instantaneous}

The second method we consider, which we call ``instantaneous relaxation approximation'', is based on the assumption that the ion's motion is much slower than the characteristic relaxation time $\tau_r$ of the internal dynamics. 

For each fixed value $v$ of the velocity, i.e. with fixed Doppler-shifted detunings, we obtain the fluorescence as proportional to the steady-state population of the excited level with the replacement $\Delta_j\to\delta_j(v)$. We approximate the ion's motional state as a classical thermal state, so that $v$ is a random variable that follows a Maxwell-Boltzmann distribution with temperature $T$. The total fluorescence as a function of the temperature and laser detunings is approximated by 
\begin{multline}
    \mathcal{F}(T)  = F_0\, Z^{-1} \int \left(\frac{m}{2 \pi k_B T}\right)^{1/2}  \rho_{11}[\delta_0(v),\delta_2(v)] \\
    \times\exp\left[-\frac{1}{2} \left(\frac{v}{\sigma_T}\right)^2  \right] dv \,,
    \label{ec: fluorescencia boltzmann}
\end{multline}
with $\sigma_T = \sqrt{{k_B T }/{m}}$ the thermal dispersion of the velocity. 
The resulting curve is plotted in Fig.~\ref{fig:three levels} in light blue.

In practice, since the ion is trapped, its velocity will be oscillating and thus the previous expression will only be a good approximation as long as the velocity changes slowly enough with respect to the electronic dynamics.  In fact, as seen in Fig.~\ref{fig:three levels}, this approximation holds well for low trap frequencies and shows some deviations from the benchmark method when the trap frequency is comparable to or larger than the dark-resonance width. 

More precisely, we expect that  Eq.~(\ref{ec: fluorescencia boltzmann}) describes well the spectrum of an ion when the rate at which the equilibrium population of the excited state changes is significantly slower than the relaxation rate $\gamma_r$ of the electronic dynamics at zero temperature. This can be expressed as the set of parameters for which the inequality
\begin{equation}
    \gamma_r \gg \frac{1}{\rho_{11}} \frac{\partial \rho_{11}}{\partial t} = \frac{1}{\rho_{11}}  \frac{\partial \rho_{11}}{\partial v} \frac{dv}{dt}\,
    \label{eq: relaxation 3 levels}
\end{equation}
holds at all times. Here, $\rho_{11}$ is taken to be the asymptotic excited state population corresponding to a given instantaneous velocity $v$, which in turn oscillates with the trap frequency. 

Even for a simple three-level system, assessing the validity of condition \eqref{eq: relaxation 3 levels} can be nontrivial. This is due to the presence of the dark resonance, which has an impact on both sides of the inequality: on the left, the relaxation rate can be much smaller than the scale of spontaneous decay from the excited level; on the right, the dependence of $\rho_{11}$ on the instantaneous velocity will be sharper close to the resonance. 

The most straightforward method to check the condition \eqref{eq: relaxation 3 levels} is to evaluate it numerically. In general, we find that the instantaneous approximation is not well justified for systems with pronounced dark resonances, as discussed in Appendix \ref{sec:instantaneous}.

Finally, it is interesting to note that for the simpler case of a two-level dipole transition, the instantaneous approximation typically holds. This result is derived analytically in Appendix \ref{sec:two-level}.

\subsection{Sidebands approximation}

When the characteristic time scales of the external and internal dynamics are comparable, the instantaneous approximation breaks. In this regime, the electronic degrees of freedom are influenced by the atomic velocity at earlier times. The time dependence of the motion can then be included in a Floquet-type expansion. Here we follow the approach originally introduced to describe micromotion echoes in~\cite{oberst1999resonance}, but adapted to describe secular motion. 
This procedure is closely related to the calculations presented in~\cite{sikorsky2017doppler, nunez2024dark}.

For each simulation run, we assume that the ion oscillates at the relevant secular frequency $\omega$ with an amplitude $A$ as in Eq.~\eqref{eq: ion motion}. The superoperator generating the time evolution of the electronic degrees of freedom in Eq.~\eqref{ec: ecuacion maestra} will then have time-dependent coefficients, becoming $\mathcal{M}' = \mathcal{M} + 2\Delta \mathcal{M} \cos(\omega t) $, where $\Delta \mathcal{M}$ is proportional to the oscillation amplitude $A$. The equation for the steady state can be solved by proposing solutions of the form
\begin{equation}
\rho(t) =\sum_{n=-\infty}^\infty \rho^{(n)} e^{in\omega t}
\end{equation}
with components oscillating at all integer multiples of the driving frequency.

For the calculation of fluorescence, we are only interested in the time-averaged density matrix, $\bar \rho$, which is equal to $\rho^{(0)}$. However, the evolution of each of the components is coupled with the rest. Indeed, replacing the expression above in the master equation \eqref{ec: ecuacion maestra} one obtains
 \begin{equation}
         \sum_{n=-\infty}^\infty \rho^{(n)} i n \omega e^{in\omega t} = (\mathcal{M} + 2\,\Delta \mathcal{M} \cos{\omega t})\sum_{n=-\infty}^\infty \rho^{(n)} e^{in\omega t}  \,.
 \end{equation}
From this equation a recurrence relation can be derived for $\rho^{(n)}$. In order to solve this relation and find the relevant contribution $\rho^{(0)}$ one resorts to a truncation \cite{oberst1999resonance}. This is justified because the amplitude of each sideband (labeled by $n$) scales approximately as the corresponding Bessel function evaluated in the modulation factor $J_n(k A)$, with $k$ the largest wavevector. One can then truncate the expansion, setting to zero the terms above a certain $n_\text{max}$ of the order of $k A$, and recursively solve for $\rho^{(0)}$.

For each value of $A$, corresponding to a certain mechanical energy $E$, the total fluorescence is proportional to the time-averaged population of the excited state, $\bar \rho_{11} (E)$. The final spectrum for a thermal state is obtained by integrating contributions as in Eq.~\eqref{eq: energy average}. We refer to this as the ``sideband approximation''.

We find that in the one-dimensional case this approximation accurately captures the system's dynamics and reproduces all the key features of the full ``driven-motion approximation'' while significantly reducing computational cost. 
We run the model with $n_\text{max} = 10$ and obtain a spectrum in 18 seconds running on one core of the same Dell PowerEdge R720xd processor used for the driven-motion approximation. The results are plotted in Fig.~\ref{fig:three levels} (pink), in complete agreement with the driven-motion curve.

Unfortunately, the effectiveness of this method diminishes when the motion in three spatial dimensions is relevant, as it can only accurately account for one oscillating frequency. We will further discuss this for the eight-level system in Sec.~\ref{sec: 8 levels}.

\subsection{Effective dephasing approximation}

The last method we consider is based on Ref.~\cite{rossnagel2015fast}, where the temperature is added by introducing an effective thermal dephasing in the model. This approximation drastically reduces the computational cost of fitting experimental spectra, but yields results that are not always in good agreement with the actual temperature. However, as we show here, this method can be calibrated to produce accurate results. 

The strategy consists in accounting for the ion's motion through an effective dephasing of the electronic levels. This dephasing is chosen proportional to the Doppler broadening corresponding to a certain temperature $T$. Following~\cite{rossnagel2015fast}, we focus on the Doppler broadening of the two-photon transition between $\ket{0}$ and $\ket{2}$, since it has the largest impact on the dark resonance. The effective dephasing rate $\Gamma_D$ depends on the absolute value of the wavevector difference, $| \vec{k}_0 - \vec{k}_2 |$, and on the standard deviation of the thermal velocity distribution:
\begin{equation}
 \Gamma_D \propto \left.\mid \vec{k}_0 - \vec{k}_2\right.\mid \sqrt{\frac{k_B T}{m}}
 \label{eq: gammaD tres niveles}\,.
\end{equation}

Since one resorts to an effective dephasing associated with the difference between the two Doppler shifts, one must decide how to include it in \eqref{eq: jump operators}. We add $\Gamma_D$ to the dissipators corresponding to each of the linewidths as 
\begin{equation}
    \Gamma_{lj} \to \Gamma_{lj}^{\rm (eff)}=\sqrt{\Gamma_{lj}^2+\Gamma_D^2\, \frac{k_j^2}{k_0^2+k_2^2}} \,\,\,\, ,\,\,\, j = 0,2\,,
\end{equation}
i.e. weighing $\Gamma_D$ by the ratio between wavevectors. In typical experiments the laser linewidths $\Gamma_{lj}$ can be made small enough such that the above expressions are dominated by thermal effects, so that
\begin{equation}
    \Gamma_{lj}^{\rm (eff)} \simeq \Gamma_D \frac{|k_j|}{\sqrt{k_0^2+k_2^2}} \,\, ,\,\,\, j = 0,2\,.
\end{equation}

We note that our particular choice of share of thermal dephasing to each laser does not have a significant impact on the depth or shape of the dark resonances. What is critical is the choice of the prefactor in the definition of $\Gamma_D$, Eq.~\eqref{eq: gammaD tres niveles}. In \cite{rossnagel2015fast} a prefactor of $1/\sqrt{2}$ was proposed. We find in our analysis that, when compared with the driven-motion curves, the choice of this prefactor can underestimate or overestimate the temperature depending on the experimental parameters. 

Two examples of underestimation can be seen in Fig.~\ref{fig:three levels}. In blue we show the spectra generated by setting the prefactor as $1/\sqrt{2}$ and the input temperature value of 20~mK, obtaining a clear disagreement with the driven-motion approximation. In blue dashed line we display the result from fitting the driven-motion curve with the dephasing model taking the temperature as a free parameter. In this case the model reproduces the correct behavior but with a different value of temperature, of about 10~mK. 

\begin{figure}[t]
    \centering
    \includegraphics[width = \columnwidth]{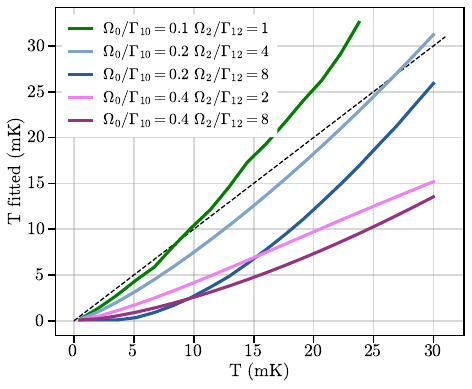}
    \caption{
    Dephasing model calibration curves, showing the value fitted by the model vs. actual temperature set for the OBS simulation. 
    Each colored curve corresponds to a different choice of Rabi frequencies as indicated in the inset. One can see how they depart from the ideal result shown as a dashed line. 
    The model is a three-level system with decay rates that match those of $^{40}$Ca$^+$ ions, {\it i.e.} $\Gamma_{10}= 2\pi\times 21.58$~MHz and $\Gamma_{12} = 2\pi\times  1.35$~MHz~\cite{hettrich2015}. The detuning is $\Delta_0 = - 2 \pi \times 15$~MHz. 
    \label{fig: temp comp 3 niveles}}
\end{figure}

To study this in more detail, we fit the spectra generated from the time-dependent Bloch equations for various temperatures by the curves obtained with effective thermal dephasing. The temperature extracted from the best fit is plotted as a function of the input temperature for different Rabi frequencies in Fig.~\ref{fig: temp comp 3 niveles}. The plot indicates that the approximation of thermal effects by effective dephasing as discussed above can be particularly poor for temperatures below 10 mK, where the fit can underestimate the input temperature by an order of magnitude. Besides, the relation between the fit result and the input temperature depends on the Rabi frequency. Nevertheless, the effective dephasing technique can provide a good temperature estimate in cases where the approximate Rabi frequencies are known, allowing for a calibration of the fitted temperature. In the following section, we assess this point again for the experimentally relevant eight-level system.

\section{Eight-level system}

\label{sec: 8 levels}

The analysis of the three-level system gives insight into how one can model the effect of thermal motion in a fluorescence spectrum exhibiting dark resonances. In particular, the characteristic relaxation rates are much lower than in the case of a dipolar transition of a two-level system, compromising the quality of the instantaneous approximation. This becomes even more problematic in the case of the eight-level system we now discuss. On the other hand, the sideband method takes into account the relaxation time, but strongly relies on the assumption that only one frequency of oscillation is relevant. The consequences of this approximation will be explored in the following.

\begin{figure}[t]
    \centering
    \includegraphics[width = 0.7\columnwidth]{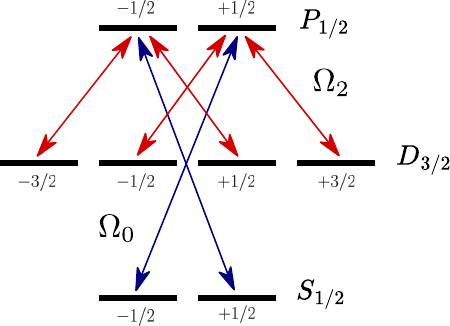}
    \caption{Energy diagram of an eight-level system corresponding to the $^{40}$Ca$^+$ ion, forming a $\Lambda$-type scheme with sublevels. Under the effect of an external magnetic field, the Zeeman shift breaks the degeneracy within each level. The polarizations of the lasers determine the allowed transitions, marked by arrows in the diagram for the case when both lasers are linearly polarized in a direction orthogonal to the magnetic field.}
    \label{fig: diagrama 8 niveles}
\end{figure}

We tackle a system corresponding to an eight-level model of $^{40}$Ca$^+$, whose energy diagram is sketched in Fig.~\ref{fig: diagrama 8 niveles}. The ground $S$ and excited $P$ states are connected by a dipolar transition driven near 397~nm with Rabi frequency $\Omega_0$ and the metastable $D$ level is depopulated by driving the $D$-$P$ transition near 866~nm with Rabi frequency $\Omega_2$. The decay rate from $P$ to $S$, $\Gamma_{10}$, is set to $2 \pi\times 21.58$~MHz, and the one from the $P$ to the $D$ manifold, $\Gamma_{12}$, to $2 \pi\times 1.35$~MHz~\cite{hettrich2015}. The ground $S$ state and excited $P$ state are doubly degenerate, while the metastable $D$ level has four-fold degeneracy. In the presence of an external magnetic field, the Zeeman splittings break the degeneracy within each manifold. The Hamiltonian and dissipator of this eight-level system are described in detail in \cite{rossnagel2015fast,NunezBarreto2022threelaser}.

Depending on the relation between the direction of the magnetic field and the polarization of the lasers, only some transitions are allowed. As in the three-level case, the formation of dark states leads to dark resonances in the spectrum. In this case, however, many more dark resonances may appear, corresponding to coherent superpositions of different sublevels of the $S$ and $D$ manifolds. The polarizations of the lasers relative to the magnetic field will determine the amount of dips that appear in the CPT spectra. We will study the case when both lasers are linearly polarized orthogonal to
the magnetic field, leading to four dark resonances.

\subsection{Comparison of simulation procedures for the one-dimensional case}

\begin{figure}[ht!]
    \centering
    \includegraphics[width = \columnwidth]{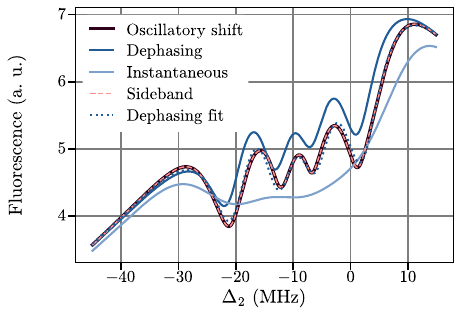}    
\caption{Comparison of eight-level spectra calculated with different methods, for $T=10$ mK and assuming $^{40}$Ca$^+$ ions. The curves from the sideband method are indistinguishable from the OBS with the oscillatory shift. The blue dotted line is the spectrum calculated by including thermal effects as dephasing, but for the temperature that yields the best fit, in this case $T_\text{fit} = 5.7$ mK. Rabi frequencies are $\Omega_0 = 0.4 \Gamma_{10}$, $\Omega_2 = 8 \Gamma_{12}$ and the detuning of the fixed laser is $\Delta_0 = - 2 \pi \times 10 $ MHz.}
\label{fig: spectra eight levels}
\end{figure}

In Fig.~\ref{fig: spectra eight levels} we show the spectra generated for the eight-level system using the different numerical approaches considered: the effective dephasing, the instantaneous approximation, the OBS with oscillatory shift, and averaging over sidebands. The results shown correspond to $T = 10$~mK. 

As with the three-level system, the temperature range within which the instantaneous approximation describes well the spectrum depends strongly on the parameters chosen. For the experimentally realistic parameters that we explore, the instantaneous approximation generally does not reproduce well the spectrum, since it drastically suppresses the dark resonances. Besides, we find that the calculation of the spectra by averaging sidebands or using the full OBS has a high computational cost for an eight-level system. Thus, it can be convenient to use the dephasing model to fit experimental data. Actually, as seen in Fig.~\ref{fig: spectra eight levels}, the comparison of the fit using the effective dephasing model (in dotted blue) with the more accurate calculation (in black) suggests that the dephasing method does convey results that resemble the true spectrum, but for a different temperature. 

\begin{figure}[t]
    \centering
    \includegraphics[width = \columnwidth]{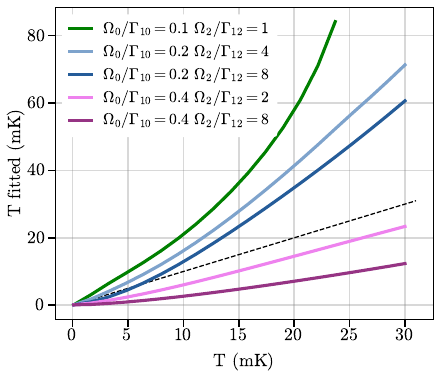}
    \caption{Temperature resulting from a fit by the dephasing model as a function of the input temperature for the spectra generated by the Sidebands approximation, for different Rabi frequencies. The ideal result is shown as a dashed line. Parameters are taken from the eight-level system corresponding to $^{40}$Ca$^+$ ions, {\it i.e.} $\Gamma_{10}= 2\pi\times 21.58$~MHz and $\Gamma_{12} = 2\pi\times  1.35$~MHz~\cite{hettrich2015}. The detuning is $\Delta_0 = - 2 \pi \times 10$~MHz.}
    \label{fig: temp comp 8 niveles}
\end{figure}

In Fig.~\ref{fig: temp comp 8 niveles} we compare the temperature that gives the best fit using the dephasing model with the input temperature for the spectrum calculated using sidebands, for different choices of the Rabi frequencies. 
As in the case of the three-level system, we find that, keeping all other input parameters fixed, the relation between both temperatures is monotonic but not always linear. We note that to properly fit the spectra, the Rabi frequencies must be left as free parameters of the dephasing model. 

With a proper estimation of the Rabi frequencies, the use of the dephasing model to measure temperatures from fits of the spectrum can be practical and convenient. A possible route to extract the Rabi frequencies is to fit a spectrum from a cold ion, near 1~mK for instance, such that thermal effects are negligible. With this information one can choose the appropriate calibration curve.

\subsection{Spectrum of an ion moving in three dimensions}

The results so far suggest that one can perform the calibration of the spectra from the simulation with the sideband method, that is computationally less demanding than the solution of the OBS with oscillatory shift. However, we have assumed that the Doppler shift could be described as a harmonic oscillation with a single frequency, which corresponds to the case when the propagation direction of both lasers coincides with one trap axis. When more motional frequencies are present, the sideband method cannot include them all, as it is based on a recursive solution with a single frequency. 

\begin{figure}[ht!]
    \centering
    \includegraphics[width=1\linewidth]{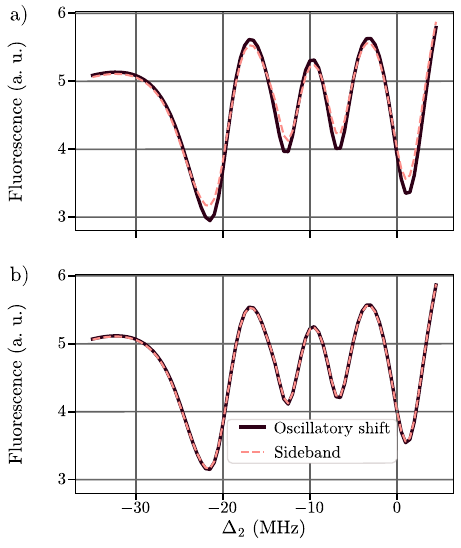}
    \caption{CPT spectrum generated by the oscillatory shift approximation with three-dimensional motion and by the sidebands approximation considering only the motion along the $z$ direction. 
    In a) the laser has comparable projections along the trap's axes, $\hat{k} \propto (1,1,2)$, whereas in (b)  the laser is mainly aligned along the trap's $z$ direction, $\hat{k} \propto (1,1,8)$. We observe a better agreement between the two methods in b), where one motional frequency dominates. 
    The parameters are, for both figures, $T = 10 $mK, $\omega_x = 2 \pi \times  0.6 $ MHz, $\omega_y =2 \pi \times 0.8  $ MHz, $\omega_z = 2 \pi \times  1.2 $ MHz, $\Omega_0 = 0.4 \Gamma_{10}$, $\Omega_2 = 8\Gamma_{12}$, $\Delta_0 = -2 \pi \times 15 $ MHz. }
    \label{fig: comparison 1d-3d}
\end{figure}

One can still approximate the spectrum of an ion moving in three directions by the sideband method considering only motion along one trap axis. To choose this axis, we take the direction with the largest projection of the driving lasers, which we assume to be collinear. As can be seen in Fig.~\ref{fig: comparison 1d-3d} a), when the effective wavevector has a comparable projection onto different axes of the trap, the sideband approximation deviates slightly from the full model. However, as expected, if one assumes that the projection of the lasers is much larger in one of the directions of the trap, the two spectra become very similar, as shown in  Fig.~\ref{fig: comparison 1d-3d} b). 

Depending on the configuration of the lasers with respect to the trap, the sideband approximation can be more or less reliable to calibrate the temperature of the ion. For example, in Fig.~\ref{fig: comparison 1d-3d} a) the fit of the full oscillatory shift curve by the dephasing model corresponds to a fit temperature of 3.4 mK whereas fitting the spectrum produced by the sideband approximation with only one of the frequencies yields a fit of 4.7 mK.  In contrast, for the case of Fig.~\ref{fig: comparison 1d-3d} b) where the projection of the laser is mainly along the $z$ trap axis the values from the fits are $T = 4.6$ mK and $T = 4.7$ mK respectively, so that the calibration becomes more accurate. In all cases, the input temperature for the fitted spectra is of 10~mK.

Finally, we note that further numerical experiments indicate that the error introduced by using the sideband approximation neglecting two of the motional frequencies decreases as the Rabi frequencies increase.

\section{Conclusions}

\label{sec:conclusions}

We studied how to estimate the temperature of a single ion through dark-resonance spectra using four different numerical approaches, all of them involving a semiclassical approximation for the motion. The most accurate techniques we evaluated are computationally very costly, while the more efficient ones produce inaccurate results. However, we showed that these faster methods can be properly calibrated, yielding a beneficial combination of accuracy and computational economy.

We consider as a benchmark the oscillatory shift approximation, where we solve the Optical Bloch Equations with the addition of time dependent detunings and average the steady state over a thermal distribution. This is  computationally the most costly approach. As alternatives, we consider the sideband approximation, a Floquet-type calculation of the steady state, and the instantaneous relaxation approximation, which neglects memory effects in the electronic dynamics. These techniques significantly reduce the computation time by avoiding the need to solve differential equations, but each comes with important limitations. The sideband approximation is accurate when one can disregard the presence of more than one motional frequency. On the other hand, the instantaneous approximation becomes incorrect when the electronic relaxation times are comparable to the motional time scales, which makes it inapplicable for the physical parameters of interest.


Of particular interest for thermometry is the dephasing method, where the Doppler broadening is approximated by an effective dephasing of the lasers. This procedure is computationally inexpensive while still reproducing important spectral features. Indeed, this idea has been used in several articles to obtain fits of the spectrum in order to estimate the temperature. Yet, our results demonstrate that, depending on the parameter regime of interest, this kind of fit can give temperature estimates that are off by up to an order of magnitude. Nonetheless, we described how this issue can be counteracted with a careful calibration of the experiment.

Our analysis focused on temperatures of the order of tens of milliKelvin, a regime in which the motion of the ion can be treated under semiclassical approximations. This choice of temperature range is related to the longer-term goal of performing reliable simulations of vibrational energy transport, thermalization, and non-equilibrium thermodynamics in ion traps. As the methods developed in this work rely on the velocity of each ion, they are well suited for the study of heat transport in ion crystals. This requires local temperature measurements which are inaccessible with other methods, such as sideband-resolved spectra, in which the populations of collective normal modes are probed. 
To this end, with the use of multi-pixel detectors, such as EMCCD cameras or structured PMTs, it would be possible to simultaneously detect fluorescence from multiple ions and extract temperature information via CPT spectra from each ion. In this way, one can measure  temperature distribution profiles in trapped ion chains.

\section*{Acknowledgements}
This work was supported by Agencia I+D+i grants PICT2018-3350, PICT2019-4349,  PICT 2020-SERIEA-00959 and PICT 2021-I-A-01288, and Universidad de Buenos Aires grant UBACyT2023-20020220400119BA. Computational resources for this work were provided by the HPC center DIRAC, funded by Instituto de Fisica de Buenos Aires (UBA-CONICET) and part of SNCAD-MinCyT initiative, Argentina.

\section*{Data availability}
The supporting data for this article are openly available from the Harvard Dataverse \cite{DVN/LS1WOY_2025}.
\bibliography{bibliography.bib}

\appendix

\section{Spectrum of a two-level system}
\label{sec:two-level}

In order to better illustrate the effect of thermal motion on the fluorescence spectrum, here we also study the simplest possible model, i.e.~the two-level atom. We consider only two electronic levels of an atom at rest, with a monochromatic wave driving the transition from the ground state $\ket{0}$ to the excited state $\ket{1}$. Under the rotating wave approximation and in the rotating frame, the dynamics are described by the Hamiltonian 
\begin{equation}
H=\hbar\left(\begin{array}{ccc}
\Delta & \frac{\Omega}{2}  \\
\frac{\Omega}{2} & 0 
\end{array}\right).
\label{ec: hamiltoniano rwa}
\end{equation}
where $\Omega$ is the Rabi frequency and $\Delta$ the detuning of the drive to the transition, with the sign convention $\Delta = \omega_l - \omega_0$ where $\omega_0$ is the atomic transition frequency and $\omega_l$ is the laser frequency. The Hamiltonian above is written in the basis $\{\ket{0}, \ket{1}\}$.

The system's master equation is: 
\begin{equation}
    \frac{d\rho}{dt} = \mathcal{M} \rho = - \frac{i}{\hbar} \left[ H, \rho \right] + \mathcal{L}_\text{damp}(\rho)
    \label{ec: ecuacion maestra 2niveles}
\end{equation}
where $M$ is a superoperator with 
\begin{equation}
\begin{split}
    &\mathcal{L}_{\text {damp }}(\rho)=-\frac{1}{2} \left[\hat{C}^{\dagger} \hat{C} \rho+\rho \hat{C}^{\dagger} \hat{C}-2 \hat{C} \rho \hat{C}^{\dagger}\right]\\
    &\hat{C}  = \sqrt{\Gamma} \,\ket{0}\bra{1},
    \label{ec: ldamp}
\end{split}
\end{equation}
where $\Gamma$ is the rate of spontaneous decay. We do not include here dephasing due to laser frequency fluctuations since they have a negligible impact in this system; this need not be the case for models including more electronic levels.
The stationary solution when the ion is at rest has an analytical expression, 

\begin{equation}
     \rho_{11} = \frac{\Omega ^2}{\Gamma ^2+4 \Delta ^2+2 \Omega ^2}\,.
     \label{ec: pexc 2level}
\end{equation}

To include the temperature of the ion in the model we follow the same four different approaches as with the three- and eight-level systems. The validity of the approximations for this system differs from the more complex ones, since the absence of dark resonances means that the characteristic time scale of the system is, at least, the decay rate of the corresponding dipolar transition. As in the main text, we take the oscillatory shift approximation as a benchmark.

\begin{figure*}[ht!]
\begin{subfigure}{\columnwidth}
    \includegraphics[width = \textwidth]{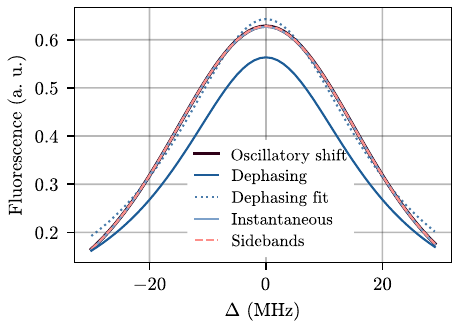}
    \caption{$\Omega/\Gamma = 0.1$}
    \label{fig:comp spect 2l low sat}
\end{subfigure}
\hspace{1mm}
\begin{subfigure}{\columnwidth}
    \includegraphics[width = \textwidth]{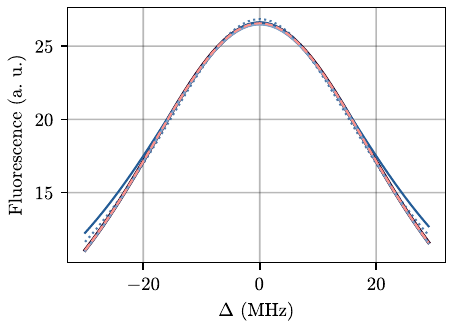}
    \caption{$\Omega/\Gamma = 1$}
    \label{fig:comp spect 2l high sat}
\end{subfigure}
\caption{Comparison of two-level spectra calculated with the four methods described, for $T=100$ mK and assuming $^{40}$Ca$^+$ ions driven at the dipolar transition at 397 nm. The Instantaneous, Sidebands and OBS with oscillatory shift curves are overlapped. The pointed line is the spectra calculated by including thermal effects as dephasing, setting the temperature that yields the best fit, $T = 119$~mK and $T = 58$~mK in a) and b), respectively.}
\label{fig: spectra two levels}
\end{figure*}

The comparison of the four methods described in the main text for a two-level system can be seen in Fig.~\ref{fig: spectra two levels}, for different values of the ratio $\Omega/\Gamma$, and taking $\Gamma = 2\pi\times 21.58$ MHz. We note that choosing a different value of $\Gamma$ is equivalent to rescaling the temperature accordingly, given the very few parameters in this simplified model. The results from oscillatory shift and instantaneous approximation are indistinguishable for the parameter ranges in the figure, which correspond to typical experimental values.  The sidebands approach is a very good approximation as long as one sums over enough sidebands. In our case, we find that by taking $n_{max} = k A$, the method gives a good result. On the contrary, the effective dephasing leads to a much worse approximation of the spectrum. We note, however, that all curves become very similar for lower temperatures.

\bigskip

\section{Validity of the instantaneous relaxation approximation}

\subsection{Two-level system}

The instantaneous relaxation approximation is, as described in Subsec.~\ref{subsec:instantaneous}, a calculation neglecting memory effects in the electronic dynamics. Under the semiclassical approximation, one can calculate the fluorescence for a thermal state as \ref{ec: pexc 2level} 
\begin{equation}
    \mathcal{F}(T,\Delta) \propto \int \left(\frac{m}{2 \pi k_B T}\right)^{1/2}  \rho_{11}(\Delta-kv) e^{-\frac{1}{2} \left(\frac{v}{\sigma_T}\right)^2  } dv \,,
    \label{ec: fluorescencia boltzmann 2d}
\end{equation}
where 
\begin{equation}
    \sigma_T = \sqrt{\frac{k_B T }{m}} \,.
    \label{eq: sigmaT}
\end{equation}
In the expression above, one can directly replace the analytical form of Eq.~\eqref{ec: pexc 2level} for the excited-state population. The shape of the resulting spectrum is called a Voigt profile~\cite{Demtroder2002-nl}, the convolution of a Lorentzian profile and a Gaussian probability density.

This approximation will be good as long as the atomic velocity changes slowly enough. More precisely, we expect that  Eq.~(\ref{ec: fluorescencia boltzmann 2d}) describes well the spectrum of an ion when the following inequality holds at all times:
\begin{equation}
   \frac{1}{\rho_{11}} \frac{\partial \rho_{11}}{\partial t}\sim\frac{1}{\rho_{11}} \frac{\partial \rho_{11}}{\partial v} \,\omega v \sim \frac{ 8 k v \omega \delta }{\Gamma^2 + 2\Omega^2 + 4\delta^2 }   \ll \frac{\Gamma}{2}
\end{equation}
where $\rho_{11}$ is taken to be the asymptotic excited state population corresponding to a given instantaneous velocity $v$ which in turn oscillates with the trap frequency. For values of detuning of the order of $\Gamma$, we have
\begin{equation}
     k v \ll \frac{5\Gamma^2}{16\omega}. 
\end{equation}
This relation takes a more complicated, but less restrictive form for smaller absolute values of the detuning. 

For $^{40}$Ca$^+$ ions, the dipole transition used for Doppler cooling and fluorescence detection has a natural linewidth given by $\Gamma \approx 2\pi\times 21.58$ MHz. Considering typical trap frequencies, for temperatures below $100$ mK the condition above is usually fulfilled and the instantaneous approximation is expected to hold.

\subsection{Three-level system}
\label{sec:instantaneous}

\begin{figure}[ht!]
    \centering
    \includegraphics[width=\linewidth]{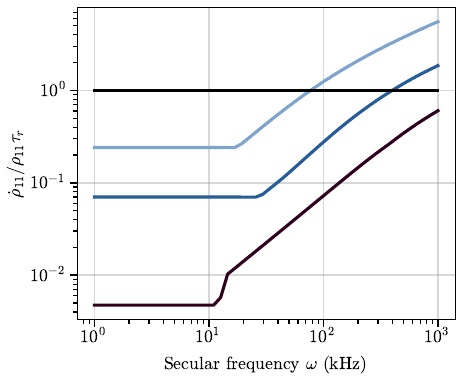}
    \caption{Relative time derivative of the excited level population divided by the relaxation rate as a function of the frequency $\omega$ of the thermal motion. We show three cases with increasing laser intensities: 
    $\Omega_0/\Gamma_{01} = \{0.1,0.3,0.5\}$ and $\Omega_2/\Gamma_{12} = \{1,3,6\}$ in colors \{light blue, blue, purple\}, correspondingly. For all plots, $\Delta_0 = \Delta_2 = - 2 \pi \times 15 $~MHz, $v = 2 \sigma_T$ with $T = 20$~mK.}
    \label{fig: inequality rate}
\end{figure}

For the three-level system, we expect the instantaneous approximation to be valid when inequality~\eqref{eq: relaxation 3 levels} holds at all times.
The relaxation rate $\gamma_r$ is calculated as the absolute value of the smallest real part of the non-zero eigenvalues of the superoperator $\mathcal{M}$ generating the evolution. 
In Fig.~\ref{fig: inequality rate} we show the quotient between the two sides of Eq.~(\ref{eq: relaxation 3 levels}) for different Rabi frequencies, for the instant at which the quotient is largest. For small secular frequencies, there is a difference of about one to two orders of magnitude for all laser intensities studied. As the frequency increases, the two sides of the inequality become comparable and the instantaneous approximation becomes less accurate relative to our benchmark. 

The figure also shows that the quality of the instantaneous approximation not only depends on temperature and trap frequency but also on the laser intensity. This is, as discussed above, a result of the fact that as the dark resonance broadens, the rate of relaxation increases.  

\end{document}